\documentclass[prl,twocolumn,showpacs,amsmath,superscriptaddress,psfig,amssymb]{revtex4}
\usepackage{graphicx}% Include figure files
\usepackage{dcolumn}% Align table columns on decimal point
\usepackage{bm}% bold math
\usepackage{multirow}

\begin{document}

%\preprint{APS/123-QED}

\title{Superconductivity in a new layered nickel-selenide CsNi$_{2}$Se$_{2}$}

\author{Huimin Chen}
\affiliation {Hangzhou Key Laboratory of Quantum Matter, Department of Physics, Hangzhou Normal University, Hangzhou 310036, China}

\author{Jinhu Yang}
\affiliation {Hangzhou Key Laboratory of Quantum Matter, Department of Physics, Hangzhou Normal University, Hangzhou 310036, China}

\author{Chao Cao}
\affiliation {Hangzhou Key Laboratory of Quantum Matter, Department of Physics, Hangzhou Normal University, Hangzhou 310036, China}

\author{Lin Li}
\affiliation {Hangzhou Key Laboratory of Quantum Matter, Department of Physics, Hangzhou Normal University, Hangzhou 310036, China}

\author{Qiping Su}
\affiliation {Hangzhou Key Laboratory of Quantum Matter, Department of Physics, Hangzhou Normal University, Hangzhou 310036, China}

\author{Bin Chen}
\affiliation {Hangzhou Key Laboratory of Quantum Matter, Department of Physics, Hangzhou Normal University, Hangzhou 310036, China}

\author{Hangdong Wang}
\email{hdwang@hznu.edu.cn}
\affiliation {Hangzhou Key Laboratory of Quantum Matter, Department of Physics, Hangzhou Normal University, Hangzhou 310036, China}
\affiliation {Department of Physics, Zhejiang University, Hangzhou 310027, China}

\author{Qianhui Mao}
\affiliation {Department of Physics, Zhejiang University, Hangzhou 310027, China}

\author{Jianhua Du}
\affiliation {Department of Physics, Zhejiang University, Hangzhou 310027, China}

\author{Minghu Fang}
\email{mhfang@zju.edu.cn}
\affiliation {Department of Physics, Zhejiang University, Hangzhou 310027, China}
\affiliation {Collaborative Innovation Center of Advanced Microstructures,  Nanjing 210093, China}

\date{\today}% It is always \today, today,

\begin{abstract}
\noindent The physical properties of CsNi$_{2}$Se$_{2}$ were characterized by electrical resistivity, magnetization and specific heat measurements. We found that the stoichiometric CsNi$_{2}$Se$_{2}$ compound undergoes a superconducting transition at \textit{T$_{c}$}=2.7K. A large Sommerfeld coefficient $\gamma$$_{n}$ ($\sim$77.90 mJ/mol$\cdot$K$^{-2}$), was obtained from the normal state electronic specific heat. However, the Kadowaki-Woods ratio of CsNi$_{2}$Se$_{2}$ was estimated to be about 0.041$\times$10$^{-5}$ $\mu\Omega$$\cdot$cm(mol$\cdot$K$^{2}$/mJ)$^{2}$, indicating the absence of strong electron-electron correlations. In the superconducting state, we found that the zero-field electronic specific heat data, $C_{es}(T)$ (0.5K $\leq$ T $<$ 2.6K), can be well fitted with a two-gap BCS model, indicating the multi-gap feature of CsNi$_{2}$Se$_{2}$. In the end, the comparison with the density functional theory (DFT) calculations suggested that the large $\gamma$$_{n}$ in these nickel-selenide superconductors may be related to the large Density of States (DOS) at the fermi surface.
\end{abstract}


\maketitle

After the discovery of the iron-based superconductors, people are exploring in the similar layered compounds where the Fe ions are replaced by the other 3d metal ions, in order to discover the next-generation of high-$\textit{T}$$_{c}$ superconductors and provide new opportunities for solid-state physics. NiPn-based (Pn: pnictogen) compounds are considered to be promising candidate for new superconductors, as the superconductivity has been observed in LaNiPO (T$_{c}$ $\sim$3 K) \cite{W.T.}, LaNiAsO (T$_{c}$ $\sim$2.75 K) \cite{L.Z.}, BaNi$_{2}$As$_{2}$ (T$_{c}$$\sim$ 0.7 K) \cite{R.F.2008}, BaNi$_{2}$P$_{2}$ (T$_{c}$ $\sim$2.4 K) \cite{T.Y.}, SrNi$_{2}$P$_{2}$ (T$_{c}$ $\sim$1.4 K)\cite{R.F.2009}, and so on. However, the superconductivity in Ni-chalcogenide is rarely reported.

Recently, KNi$_{2}$Se$_{2}$ \cite{N.J.2012} and KNi$_{2}$S$_{2}$ \cite{N.J.2013}, which have a similar structure to the iron-based superconductors A$_{y}$Fe$_{2-x}$Se$_{2}$ (A=Tl, K, Rb, Cs) \cite{G.J.,F.M.}, were reported to be superconductivity with T$_{c}$ $\sim$0.8K and $\sim$0.4K, respectively. Upon replacement of the K with Tl, the transition temperature even can rise to be 3.7K for TlNi$_{2}$Se$_{2}$ \cite{W.H.}. Compared with A$_{y}$Fe$_{2-x}$Se$_{2}$ (A=Tl, K, Rb, Cs), it is homogenous without Ni vacancy in TlNi$_{2}$Se$_{2}$, suggesting a heavy electron doping on the fermi surface. Interestingly, no superconducting transition was observed in the unstoichiometric compound K$_{0.95}$Ni$_{1.86}$Se$_{2}$ (may equal to the hole doping) down to 0.3K \cite{L.H.}, even the physical properties were very similar with that of the stoichiometric KNi$_{2}$Se$_{2}$. For these Ni-chalcogenide superconductors, the superconductivity usually appears to involve heavy electrons \cite{N.J.2012,N.J.2013,W.H.}, as inferred from the large Sommerfeld coefficient, $\gamma$$_{n}$. This result may make them be a bridge between cuprate- or iron-based and conventional heavy-fermion superconductors. By now, the origin of the large $\gamma$$_{n}$ at low temperature is not confirmed yet. An intriguing possibility is that the large $\gamma$$_{n}$ is due to the existence of local charge order, which may induce the strong electron correlations and heavy-fermion behavior \cite{N.J.2012,N.J.2013,W.H.}. However, the recent angle-resolved photoemission spectroscopy (ARPES) and optical spectroscopy studies reveal the weak correlation feature in these Ni-chalcogenide superconductors. Instead, they suggested that the large $\gamma$$_{n}$ may be ascribed to the large density states and the Van Hove singularity in the vicinity of the fermi energy \cite{F.Q.,X.N.,W.X.}.

Though the Ni-chalcogenide superconductors have been studied for several years, the superconductivity in the single crystalline sample had only been realized in TlNi$_{2}$Se$_{2}$ \cite{W.H.} and TlNi$_{2}$S$_{2}$ \cite{W.H.2013} by now. To understand the intrinsic properties, more superconducting samples, especially the alkali metal compounds, are needed to be investigated. In this paper, we synthesized the CsNi$_{2}$Se$_{2}$ single crystal successfully. The measurements of resistivity, magnetization and specific heat were carried out down to $\textit{T}$=0.5K. It was found that the stoichiometric CsNi$_{2}$Se$_{2}$ compound undergoes a superconducting transition at \textit{T$_{c}$$^{onset}$}=2.7K, which is very different from that in K$_{0.95}$Ni$_{1.86}$Se$_{2}$ single crystal \cite{L.H.}. At low temperatures, a large Sommerfeld coefficient, $\gamma$$_{n}$ ($\sim$77.90 mJ/mol K$^{2}$), was obtained from the normal state electronic specific heat. In the superconducting state, we found that the zero-field electronic specific heat data, $C_{es}(T)$ (0.5K $\leq$ T $<$ 2.6K), can be well fitted using a two-gap BCS model, indicating the multi-gap feature of CsNi$_{2}$Se$_{2}$. At the end, density functional theory (DFT) calculations were performed for TlNi$_{2}$Se$_{2}$, KNi$_{2}$Se$_{2}$ and CsNi$_{2}$Se$_{2}$, respectively, and the result was compared with their corresponding large $\gamma$$_{n}$. It suggested that the large $\gamma$$_{n}$ may be related to the large DOS at the Fermi surface in these nickel-selenide superconductors.

\begin{figure}
  \includegraphics[width=8cm]{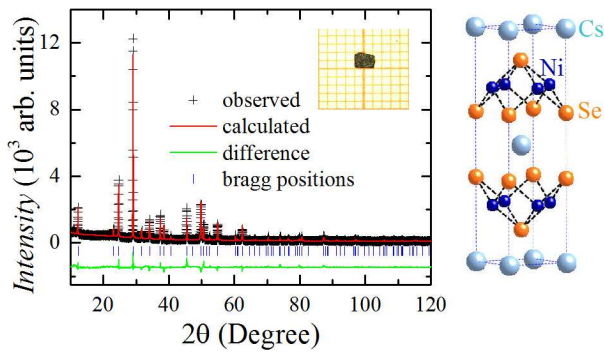}\\
  \caption{(Color online) Left: Rietveld refinement profile for the powder x-ray diffraction of CsNi$_2$Se$_2$. Right: Crystal structure of stoichiometric
CsNi$_2$Se$_2$ with the tetragonal ThCr$_2$Si$_2$-type structure. Inset:  A photo of CsNi$_2$Se$_2$ crystal.}\label{}
\end{figure}

Large plate-like single crystals of CsNi$_2$Se$_2$ were grown using self-flux method. First, the precursor Cs$_2$Se was prepared by heating Cs and Se powders at 200$^o$C. Then, Cs$_2$Se, Ni and Se powders were mixed in an appropriate stoichiometry and were put into alumina crucibles and sealed in an evacuated silica tube. The mixture was heated up to 1000$^o$C and kept over 3 hours. Then the melting mixture was cooled down to 700$^o$C in a cooling rate of 3$^o$C/h. Finally the furnace was cooled to room temperature after shutting down the power. The structure of single crystals was characterized by X-ray diffraction (XRD). To avoid exposure to air, the sample was sealed using \textit{N}-grace during the XRD data collecting. In the powder XRD patterns (figure 1), all peaks can be well indexed with a ThCr$_2$Si$_2$-type structure (space group: \textit{I}4/\textit{mmm}). The lattice parameters, \textit{a} = 3.988{\AA}, and \textit{c} = 14.419{\AA} were obtained by fitting the XRD data, indicating a larger unit cell than that of KNi$_2$Se$_2$ and TlNi$_2$Se$_2$. The elemental analysis was performed using an energy-dispersive x-ray spectroscopy (EDX) in a Zeiss Supra 55 scanning electron microscope. The EDX results indicate that the crystals are rather homogenous and the determined average atomic ratios are Cs:Ni:Se = 1.02:2.03:2.00 when fixing Se stoichiometry to be 2, confirming the stoichiometry of CsNi$_2$Se$_2$. The ab-plane $\rho$(T) measurements were performed by the standard four-probe technique using the \textit{Quantum Design} Physical Properties Measurement System PPMS-9. Temperatures down to 0.5 K were obtained using a $^{3}$He attachment to the PPMS. The heat capacity measurements were carried out using the relaxation method. The magnetic susceptibility was measured using the \textit{Quantum Design} MPMS-SQUID.

\begin{figure*}
  \includegraphics[width=13cm]{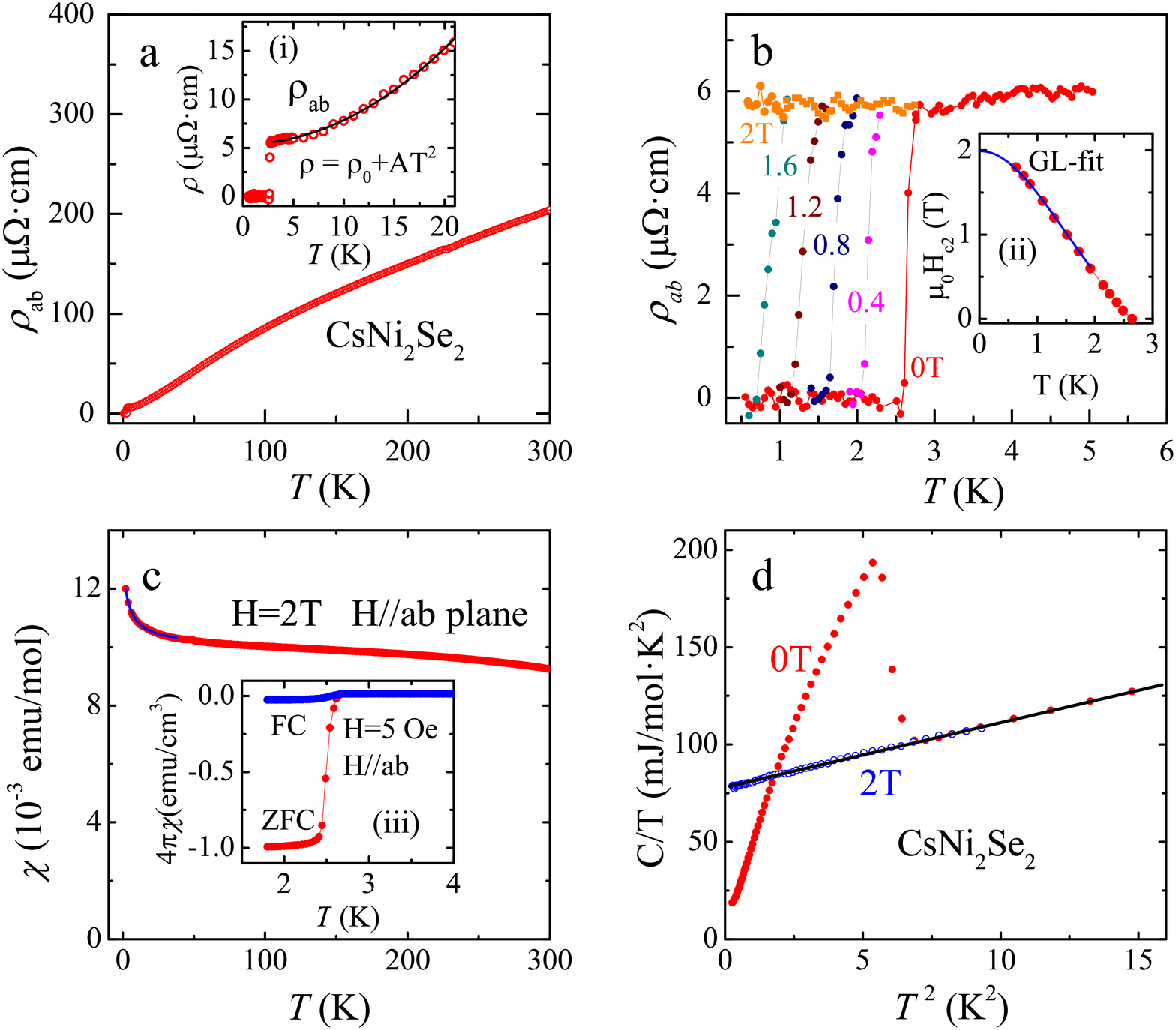}\\
  \caption{(Color online) (a) Temperature dependence of \textit{in}-plane resistivity, $\rho$$_{ab}$(T) for CsNi$_2$Se$_2$ crystal. (b) The evolution of the in-plane resistivity $\rho$$_{ab}$(T) under magnetic fields up to 2.0 T applied parallel to the \textit{c} axis. (c) Temperature dependence of the normal state magnetic susceptibility, $\chi$(T), measured at 2 Tesla field parallel to \textit{ab} plane. (d) Specific heat divided by temperature, C/T $\textit{vs.}$ T$^{2}$, measured under 0 and 2 T fields. The solid straight line is a guide to the eye. Inset: (i) $\rho$$_{ab}$(T) near the superconducting transition; (ii) Upper critical field H$_{c2}$ $\textit{vs.}$ T for CsNi$_2$Se$_2$ crystal. (iii) $\chi$(T) near the superconducting transition, measured at 5 Oe field parallel to \textit{ab} plane (for minimizing the demagnetization factor) with both zero-field cooling (ZFC) and field cooling (FC) processes.}\label{}
\end{figure*}

Figure 2(a) shows the electrical resistivity in the \textit{ab}-plane, $\rho$$_{ab}$(T), as a function of temperature for CsNi$_2$Se$_2$ crystal. The value of $\rho$$_{ab}$(300K) and $\rho$$_{c}$(300K) (not shown here) is about 204$\mu\Omega\cdot$cm and 572 $\mu\Omega\cdot$cm, respectively. Then the $\rho_c$/$\rho_{ab}$ is calculated to be 2.8, indicating the anisotropy is not so large, although CsNi$_2$Se$_2$ owns a layered structure. Upon cooling down from room temperature, $\rho$$_{ab}$(T) exhibits a metallic behavior. In the normal state, no abnormal change in $\rho_{ab}(T)$ was observed, which occurs in both iso-structural BaNi$_2$As$_2$ \cite{R.F.2008} and SrNi$_2$P$_2$ \cite{R.F.2009} compounds, corresponding to the structural transition from a tetragonal at higher temperatures to a triclinic at lower temperatures. Below $T_c$ $\sim$ 2.7K, the $\rho_{ab}(T)$ drops to zero abruptly, suggesting a superconducting transition happens. It is also confirmed by a large diamagnetic signal [see the inset (iii) of figure 2] and a specific heat jump at $T_c$ as shown in figure 2(d). The residual resistivity ratio [RRR =$\rho_{ab}$(300K)/$\rho_{ab}$(3K) $\sim$37] and superconducting transition width $\bigtriangleup$$T_c$ $\sim$0.1 K reflects the high quality of the single crystals. At low temperatures (3K $\leq$ T $\leq$ 20K), it was found that the $\rho_{ab}(T)$ can be well fitted using the the equation $\rho_{ab}(T)$=$\rho_0$+$AT^2$, where $\rho_0$ =5.368 $\mu\Omega$$\cdot$cm and \textit{A} =2.474$\times$10$^{-3}$ $\mu\Omega$$\cdot$cm/K$^2$, suggesting a Fermi liquid ground state. Combining the value of Sommerfeld coefficient $\gamma_{n}$ below, the Kadowaki-Woods ratio, A/$\gamma$$^{2}$, was estimated to be 0.041$\times$10$^{-5}$ $\mu\Omega$$\cdot$cm(mol$\cdot$K$^{2}$/mJ)$^{2}$. This value is one order of magnitude smaller than that for the standard heavy-fermion systems ($\sim$10$^{-5}$ $\mu\Omega$$\cdot$cm(mol$\cdot$K$^{2}$/mJ)$^{2}$), and is quite similar with that for many transition metals ($\sim$0.04$\times$10$^{-5}$ $\mu\Omega$$\cdot$cm(mol$\cdot$K$^{2}$/mJ)$^{2}$) \cite{K.K.,J.A.}. We suggest that this result may indicate the absence of strong correlation in this compound, which is consistent with the previous results \cite{F.Q.,X.N.,W.X.}.

Figure 2(b) is the field dependence of the $\rho$$_{ab}$(T) of CsNi$_2$Se$_2$ at low temperatures, when the magnetic field was applied parallel to the \textit{c} axis. The transition temperature $T_c$ shifts to lower temperature in external magnetic fields. Using the middle superconducting transition temperature in $\rho$$_{ab}$(T), the upper critical field $\textit{H}$$_{c2}$ is plotted as a function of temperature in the inset (ii) of figure 2(b). According to the Ginzburg-Landau theory, the zero temperature upper critical field $\textit{H}$$_{c2}$(0) can be estimated by using the formula $H_{c2}(T)= H_{c2}(0)(1-t^2)/(1+t^2)$, where \textit{t} is the reduced temperature $t=T/T_c$. The fitting result yields the value of $\mu_0H_{c2}$(0)= 2 Tesla, which is 2.5 times that of TlNi$_2$Se$_2$ \cite{W.H.}. The superconducting coherence length $\xi$$_{0}$ can be estimated from the relation $\xi$$_{0}$=[$\Phi$$_{0}$/2$\pi$H$_{c2}$]$^{0.5}$, yielding $\xi$$_{0}$$^{ab}$=12.8 nm. The normal state magnetic susceptibility, $\chi_{ab}$(T), for CsNi$_2$Se$_2$ crystal, is plotted as a function of temperature in figure 2(c). It seems nearly temperature independent at high temperatures, consistent with previous reports of Pauli paramagnetism \cite{H.G.}. At low temperatures, it exhibits a Curie-tail like behavior, which may be due to the existence of magnetic impurity (as shown as the fitting blue line in figure 2(b) corresponding to $<$ 1.00 mol $\%$ of an $\textit{S}$=1 impurity, \textit{e.g.} Ni$^{2+}$). However, we can't exclude the effect of spin fluctuation at low temperatures, which may play an important role in the superconductivity.

\begin{figure}
  \includegraphics[width=7cm]{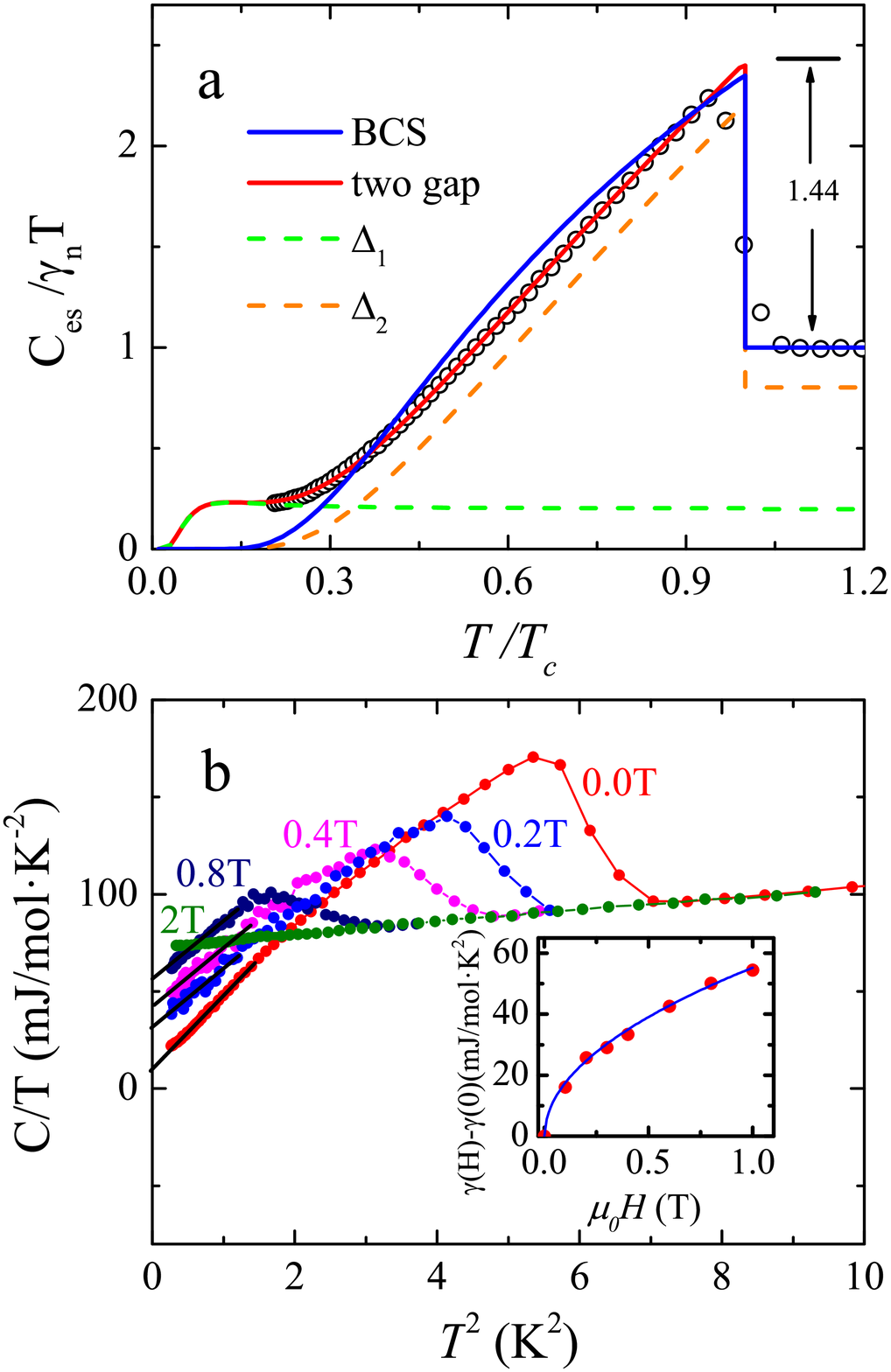}\\
  \caption{(Color online) (a) Reduced temperature $T/T_{c}$ dependence of electronic specific heat divided by temperature $C_{es}/$$\gamma$$T$ in the superconducting state at zero field, where $C_{es} = C - C_{latt}$. The two solid lines show the fitting curve of the conventional BCS model and the two-gap model to $C_{es}/$$\gamma$$T$, respectively. The two dashed lines show the individual contributions of the two gaps to $C_{es}/$$\gamma$$T$. (b) Low-temperature specific heat divided by temperature $C/T vs. T^{2}$, measured at various fields near superconducting transition. Inset: Magnetic field dependence of field-induced electronic specific-heat coefficient $\Delta$$\gamma$(H) in the mixed state.}\label{}
\end{figure}

In the nickel-chalcogenide superconductors, the superconducting transitions always accompany a large Sommerfeld coefficient $\gamma$$_{n}$ at low temperature, which may imply the non-trivial pairing mechanism. So we carried out detailed measurements of the specific heat $\textit{C}$(T,H) for CsNi$_2$Se$_2$ crystal. Figure 2(d) shows the $\textit{C}$(T,H)/$\textit{T}$ as a function of $\textit{T}$$^{2}$, measured under 0T, and 2T field, respectively. For the normal state specific heat $\textit{C}$$_{N}$(T), a linear fit of the data from 0.5K to 8K to $C/T$= $\gamma$+$\beta$$T^{2}$ gives $\gamma$=77.90 mJ/mol$\cdot$K$^{-2}$ and $\beta$=3.32 mJ/mol$\cdot$K$^{-4}$, which implies a Debye temperature $\Theta$$_{D}$ of 139K. Compared with the value of KNi$_2$Se$_2$($\sim$44mJ/mol$\cdot$K$^{-2}$) \cite{N.J.2012} and TlNi$_2$Se$_2$ ($\sim$40mJ/mol$\cdot$ K$^{-2}$) \cite{W.H.}, the Sommerfeld coefficient $\gamma$$_{n}$ for CsNi$_2$Se$_2$ is much larger. It can even compared with that for KFe$_2$As$_2$ ($\sim$94mJ/mol$\cdot$K$^{-2}$), which is widely accepted as an unconventional superconductor \cite{A.M.}.

The electronic specific heat ($C_{es}(T)$) in the superconducting state is obtained by the deduction of phonon contribution ($\beta$T$^{3}$) from the total $\textit{C}$(T). As shown in the figure 3(a), the normalized specific heat jump at $T_{c}$ ($\Delta C/(\gamma_{n}T_{c})$) is about 1.44, consistent with the theoretical value (1.43) of the well-known BCS theory. Then, we analyze the data by fitting $C_{es}(T)$ with different gap functions. First, we consider the case of a single gap $\Delta_{0}$. The temperature dependence is taken to be the same as in the BCS theory, $\textit{i.e.}$ $\Delta(t) = \Delta_{0}\delta(t)$, where $\delta(t)$ is the normalized BCS gap at the reduced temperature $t = T/T_{c}$ as tabulated by M$\ddot{u}$hlschlegel \cite{M.B.}. For a standard BCS-type superconductor, the thermodynamic properties, entropy (S) and $\textit{C}$$_{es}$, can be written as:

\begin{equation}\label{}
S =-\frac{6\gamma_{n}}{\pi^{2}} \frac{\Delta_{0}}{k_{B}}\int_{0}^{\infty}[f\ln{f}+(1-f)\ln{(1-f)}] \textrm{d} y
\end{equation}
\begin{equation}\label{}
C_{es}=T\frac{\textrm{d}S}{\textrm{d}T}
\end{equation}

where $f=[\exp[\beta E]+1]^{-1}$ and $\beta = (k_{B}T)^{-1}$. The energy of the quasi-particles is given by $E = [\varepsilon^{2}+\Delta^{2}(t)]^{0.5}$, where $\varepsilon$ is the energy of the normal electrons relative to the Fermi surface. The integration variable is $y = \varepsilon/\Delta_{0}$. Considering the multi-band feature of these nickel-chalcogenide superconductors as indicated in \cite{F.Q.,X.N.,L.F.}, we also carried out the fitting using two gap model, where the total specific heat can be considered as the sum of the contributions of each band calculated independently according to equations (1) and (2). It seems that the single-gap model fitting is not very good, especially at low temperatures, while the two-gap model presents the best fit to the $C_{es}$/\textit{T }data [see figure 3a], just as described in \cite{W.H.}. In the figure 3a, we plot the contributions from the two superconducting gaps, $\triangle_1$ = 0.22 $k_BT_c$ and $\triangle_2$ = 1.97 $k_BT_c$, as well as their sum (red line). The weight contributed from the first gap, $\triangle_1$, is about 0.20.

To get more information of the superconducting gap, we also carried out the low temperature specific heat measurements under various magnetic fields, as shown in the figure 3b. At zero field, the linear extrapolation of $C/T$ $\textit{vs.}$ $T$$^{2}$ to $T$=0K gives a 'residual' Sommerfeld coefficient of $\gamma$$_{0}$ = 5.5 mJ/mol$\cdot$ K$^{-2}$. Considering the air sensitive nature of CsNi$_2$Se$_2$, we may ascribe it to the existence of small fraction ( $\sim$6.8\%) of non-superconducting phase. With increasing of the magnetic field, the magnitude of the specific heat jump at $\textit{T}$$_{c}$ decreases, and the linear electronic specific heat coefficient, $\gamma$(H), increases. In this study, the field dependence of [$\gamma$(H)-$\gamma$(0)] obeys the Volovik relation ($\textit{i.e.}$, $\propto$ H$^{1/2}$) very well (the inset), just as observed in TlNi$_2$Se$_2$. This behavior was once considered as a common feature of the d-wave superconductors, where the H$^{1/2}$ term arises from a Doppler shift of the quasi-particle excitation spectrum in the outer regions of the vortices. Yet, this behavior may not be always so unique, because it was also observed in other s-wave superconductors, such as NbSe$_{2}$ \cite{S.D.,S.J.}, V$_{3}$Si \cite{R.A.}, and CeRu$_{2}$ \cite{H.M.}. Recently, low-temperature thermal conductivity measurements have identified the multiple nodeless superconducting gaps in TlNi$_2$Se$_2$ \cite{H.X.}, so we suggest that the CsNi$_2$Se$_2$ may have the similar properties.

\begin{figure}
  \includegraphics[width=7cm]{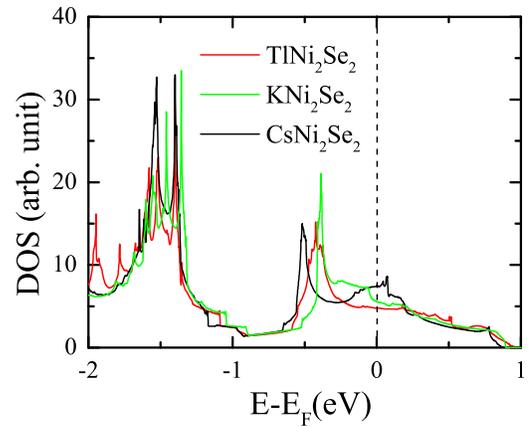}\\
  \caption{(Color online) Partial DOS of KNi$_2$Se$_2$, TlNi$_2$Se$_2$, and CsNi$_2$Se$_2$ in the -2 to 1eV energy range relative to the Fermi energy. The dashed line is a guide to eyes.}\label{}
\end{figure}

In the end, we present a roughly discussion about the origin of the large $\gamma$$_{n}$, by comparing TlNi$_2$Se$_2$, KNi$_2$Se$_2$ and CsNi$_2$Se$_2$ as typical cases, since they have the similar structure and almost same ions except the alkali metal (or thallium) ion. As mentioned above, the Sommerfeld coefficient $\gamma$$_{n}$ for TlNi$_2$Se$_2$, KNi$_2$Se$_2$ and CsNi$_2$Se$_2$ are about 40, 44, and 79 mJ/mol$\cdot$K$^{-2}$, respectively. Usually, the large $\gamma$$_{n}$ is related to the strong electron correlation. However, the ARPES and optical results have demonstrated the weak correlation in KNi$_2$Se$_2$ and TlNi$_2$Se$_2$ \cite{F.Q.,X.N.,W.X.}. As well known, the $\gamma$$_{n}$ is proportional to the DOS at $\textit{E}$$_{F}$ in a free electron framework. So we performed the DFT calculations for these three compounds (figure 4) to examine the relation between their DOS at $\textit{E}$$_{F}$ and $\gamma$$_{n}$. As shown in the figure, the DOS of all the three compounds exhibit a similar dispersion, indicating that they may contain similar electronic structure. At fermi energy, the DOS value increases in the order of TlNi$_2$Se$_2$, KNi$_2$Se$_2$ and CsNi$_2$Se$_2$. Furthermore, the estimated ratio of the DOS value in TlNi$_2$Se$_2$, KNi$_2$Se$_2$, and CsNi2Se2 is about 1 : 1.16 : 1.7, which is generally consistent with the corresponding ratio (1 : 1.1 : 1.95) of their $\gamma$$_{n}$. Therefore, we suggest that the large $\gamma$$_{n}$ may be related to the large DOS at the Fermi surface in these nickel-selenide superconductors, which has also been proposed in \cite{F.Q.} from another point of view.

In summary, we synthesized the CsNi$_{2}$Se$_{2}$ single crystal successfully. The measurements of resistivity, magnetization and specific heat were carried out down to $\textit{T}$=0.5K. It was found that the stoichiometric CsNi$_{2}$Se$_{2}$ compound undergoes a superconducting transition at \textit{T$_{c}$$^{onset}$}=2.7K. A large Sommerfeld coefficient, $\gamma$$_{n}$ ($\sim$77.90 mJ/mol$\cdot$K$^{-2}$), was obtained from the normal state electronic specific heat. However, the Kadowaki-Woods ratio of CsNi$_{2}$Se$_{2}$ was estimated to be about 0.041$\times$10$^{-5}$ $\mu\Omega$$\cdot$cm(mol$\cdot$K$^{2}$/mJ)$^{2}$, comparable with those for transition metals, which may indicate the absence of strong electron-electron correlation. In the superconducting state, we found that the zero-field electronic specific heat data, $C_{es}(T)$ (0.5K $\leq$ T $<$ 2.6K), can be well fitted with a two-gap BCS model, indicating the multi-gap feature of CsNi$_{2}$Se$_{2}$. At the end, density functional theory calculations were performed for TlNi$_{2}$Se$_{2}$, KNi$_{2}$Se$_{2}$ and CsNi$_{2}$Se$_{2}$, respectively. The result indicated that the ratio of the DOS value at E$_{F}$ is generally consistent with the corresponding ratio of their $\gamma$$_{n}$, suggesting that the large $\gamma$$_{n}$ may be related to the large DOS at the Fermi surface.

\acknowledgments
This work is supported by the National Basic Research Program of China (973 Program) under grant No. 2011CBA00103, 2012CB821404 and 2009CB929104, the Nature Science Foundation of China (Grant No. 10974175, 10934005, 11204059 and 11274006), and Zhejiang Provincial Natural Science Foundation of China (Grant No. LR12A04003), and the Fundamental Research Funds for the Central Universities of China.

\end{document}